\documentclass[aps,prb,twocolumn,floatfix,showpacs]{revtex4-1}
\usepackage{graphicx}
\usepackage{setspace}
\usepackage{rotating}
\usepackage{amsmath}
\usepackage{amssymb}
\usepackage{longtable}

\def\d{{\rm d}}

\def\vxc{{v_{\rm xc}}}
\def\vx{{v_{\rm x}}}
\def\rr{{\rm r}}
\begin{document}

\title{Kohn-Sham band gaps and potentials of solids from the optimised effective potential method within the random phase approximation}
\author{Ji\v{r}\'\i\ Klime\v{s}}
\email{jiri.klimes@univie.ac.at}
\author{Georg Kresse}
\affiliation{Faculty of Physics and Center for Computational Materials Science, University of Vienna,
Sensengasse 8/12, A-1090 Wien, Austria}

\pacs{
71.15.Mb  
71.20.-b 
}

\date{\today}
\begin{abstract}

We present an implementation of the optimised effective potential (OEP) scheme for the 
exact-exchange (EXX) and random phase approximation (RPA) energy functionals 
and apply these methods to a range of bulk materials.
We calculate the Kohn-Sham (KS) potentials and the corresponding band gaps and compare them to the potentials
obtained by standard local density approximation (LDA) calculations.
The KS gaps increase upon going from the LDA to the OEP in the RPA and finally to the OEP for EXX.
This can be explained by the different depth of the potentials in the bonding and interstitial regions.
To obtain the true quasi-particle gaps the derivative discontinuities or $G_0W_0$  corrections need to be added to the  RPA-OEP KS gaps. 
The predicted $G_0W_0$@RPA-OEP quasi-particle gaps are about 5\% too large compared to the experimental values.
However, compared to $G_0W_0$ calculations based on local or semi-local functionals, where the errors vary between different materials, 
we obtain a rather consistent description among all the materials.

\vskip 1cm
Copyright (2014) American Institute of Physics. This article may be
downloaded for personal use only. Any other use requires prior permission of the author
and the American Institute of Physics.
The following article appeared in J. Chem. Phys. {\bf 140}, 054516 (2014) and may be found at
\href{http://scitation.aip.org/content/aip/journal/jcp/140/5/10.1063/1.4863502}{http://scitation.aip.org/content/aip/journal/jcp/140/5/10.1063/1.4863502}.

\end{abstract}

\maketitle

\section{Introduction}


Kohn-Sham (KS) density functional theory (DFT) has developed into a widely used scheme
for the description and prediction of materials properties.
The central quantity in KS DFT is the functional for the exchange-correlation (XC) energy 
($E_{\rm xc}$) which includes all the electronic interactions beyond the Hartree term.
It is now a well established fact that even simple approximations for $E_{\rm xc}$ based on the local electron density alone 
often give sufficiently accurate results and this makes DFT so successful.
However, functionals that are based solely on the density, e.g., the local density approximation (LDA), 
or the generalized gradient approximations (GGA), cannot be easily systematically improved.
In fact, most of the recent development has focused on functionals that include the KS orbitals and
possibly their energies in the energy functional, for example meta-GGAs and hybrids.
A family of functionals that offers a route for systematic improvements are functionals based
on perturbation theory, such as the many-body perturbation theory (MBPT).\cite{casida1995,barth2005}
Here the accuracy is increased by taking progressively more terms in the perturbation series.
The downside of such functionals is that compared to the standard schemes the cost to evaluate
the energy is higher.

The second central quantity in the KS DFT scheme is the XC potential $\vxc$, defined as 
$\vxc(\rr)=\delta E_{\rm xc}/\delta \varrho(\rr)$.
The XC potential is required if one wants to perform self-consistent calculations,
and it can be obtained rather straightforwardly for the explicit functionals of the density.
However, it is more involved to obtain the potential for orbital dependent functionals
and $\delta E_{\rm xc}/\delta \varrho(\rr)$, performed with chain derivatives, leads to the 
so-called optimized effective potential (OEP) equation.\cite{footGKS}
In the case when only the exchange interaction is included, corresponding to the Hartree-Fock (HF) method,
the OEP equation was obtained by Sharp and Horton already in 1953\cite{sharp1953} following the work
of Slater\cite{slater1951} who sought to find a common local potential within the HF method.
The equations have been given later by other authors, who also realised that this common local
potential is the KS potential in the exchange only case.\cite{talman1976,sahni1982,gorling1994,gorling1996}
Furthermore, this scheme, usually denoted as exact-exchange OEP (EXX-OEP) or exchange-only OEP,
has been implemented for calculations of atoms, molecules, 
and solids.\cite{kotani1997,stadele1997,gorling1999,ivanov1999,stadele1999,aulbur2000,fleszar2001,engel2001,magyar2004,engel2009prl,betzinger2011,betzinger2012}
As the correct asymptotic behavior ($-1/r$) of the potential is obtained for finite systems,
properties of interest, such as electron affinities or ionisation potentials, tend to be 
improved.\cite{gorling1999}
Making approximations to the EXX-OEP equations has also received considerable interest.
Well known is the Krieger-Li-Iafrate (KLI) approximation\cite{krieger1992p1,krieger1992p2} and similar 
approaches,\cite{sala2001,gruning2002,iafrate2013,ryabinkin2013}  applied to both molecular and condensed 
matter systems.\cite{kleinman1994,bylander1995prb,bylander1995,bylander1996}

There has been considerable interest to go beyond the exchange only case, mostly when describing 
condensed matter systems where screening is important.
Within MBPT this is done by including progressively more electron interaction terms in the perturbative series.
A well known example is the $GW$ approximation for the electron self-energy, where the electrons interact via 
a screened Coulomb interaction $W$. This method is widely used to obtain quasi-particle energies.
However, MBPT can be used to obtain total energies as well.\cite{klein1961,luttinger1960}
The MBPT energy functionals have received some interest recently, and there are several publications that discuss in detail 
the particular choice of the MBPT functional and self-energy approximation and how these relate
to KS DFT.\cite{sham1985,niquet2003pra,niquet2004,barth2005,gruning2006,hellgren2007}
In fact, the XC functional of the Kohn-Sham DFT can be simply written using the MBPT expression for the
electron-electron interactions.\cite{barth2005,casida1995,baym1961}
For example, when the $GW$ self-energy is inserted into the MBPT energy functional of Klein,\cite{klein1961}
the so-called random phase approximation (RPA) energy formula is obtained. 
More specifically, the direct RPA is obtained, which is usually derived from the adiabatic-connection fluctuation-dissipation theorem.\cite{langreth1975,langreth1977,gunnarsson1976,dobson1996,furche2001}
The OEP equation for this case, and the general case when the self energy is non-local and energy dependent,
was derived by Sham and Schl\"{u}ter who used the fact that the KS and interacting densities should be
identical.\cite{sham1983}
Thus within the MBPT formalism the OEP procedure for the RPA energy functional involves the $GW$ self-energy,
and in the following, we simply refer to this procedure as RPA-OEP.
Because of the complexity of the RPA-OEP scheme, it has been applied only to few solid state systems so far,\cite{godby1988,eguiluz1992,gruning2006}
sometimes with approximations made to the correlated part of self-energy.\cite{kotani1998mmm,kotani1998jpcm}
There have been also numerous applications of OEP within RPA or MP2 energy functionals
to atoms or molecules.\cite{bonetti2001,grabowski2002,niquet2003,niquet2003comm,bonetti2003comm,hellgren2007,hellgren2008,hellgren2010,hellgren2012,verma2012,bleiziffer2013}
In some of the publications, different derivations of the OEP equation were given that-- unlike the one of
Sham and Schl\"{u}ter --do not involve the electronic self-energy but rather rely on
total energy expressions\cite{bonetti2001,bleiziffer2013} or expressions for the electron density.\cite{verma2012}

One of the fundamental questions concerning the OEP KS scheme is the magnitude of the true KS gap.
It is known that the experimental quasi-particle gap does not correspond to the KS single 
particle gap since there is a derivative discontinuity in the potential (for the orbital dependent 
functionals).\cite{perdew1982,perdew1983,sham1983}
Hence to obtain the quasi-particle gap one usually resorts to MBPT and performs, e.g., $GW$ calculations.
However, for semiconductors the EXX-OEP gaps are surprisingly close to the experimental quasi-particle gaps,
and one was thus lead to believe that there is a physical reason for this.\cite{stadele1997,stadele1999}
However, the agreement diminishes for systems with a large gap.\cite{magyar2004}
Furthermore, since only the exchange diagram is used in EXX, the calculations
will yield basically HF single particle energies once the derivative discontinuity is added,
and the HF gap is known to be too large.
Gaps closer to the experiment are obtained for the RPA-OEP scheme with the derivative discontinuity 
added.\cite{godby1988,gruning2006}
Interestingly, it was found that the single particle RPA-OEP gaps are rather close to LDA gaps
but this point has not been settled upon yet and to study if and why these two gaps are close
is one of the objectives of this study.

In this work we study the KS potentials and gaps as obtained with the EXX-OEP and RPA-OEP schemes.
In Section~\ref{sec_gaps} we discuss the band gaps and compare to previous EXX-OEP and RPA-OEP calculations.
Interestingly, the RPA-OEP and LDA gaps do not agree that closely. The RPA-OEP one particle gaps are generally larger by several tenths 
of an eV for semiconductors.
A similar increase in the one particle gap is observed upon going from RPA-OEP to EXX-OEP.
In some cases, though, there are large differences between LDA and RPA-OEP, for example, the RPA-OEP gap is almost 
1~eV larger than the LDA gap for ZnO.
Furthermore, we study the differences between the EXX-OEP, RPA-OEP, and LDA potentials in Section~\ref{sec_pot}.
We show that there are two main factors that lead to the differences in the gaps.
First, the proper treatment of self-interaction in EXX-OEP or RPA-OEP deepens the potentials
in the core and bonding regions compared to LDA.
Second, in the interstitial regions the EXX-OEP is more repulsive than the RPA-OEP;
this can be attributed to the inclusion of response of the electron density in the $GW$ self-energy.
Overall, this leads to the single-particle KS gaps following the order LDA$<$RPA-OEP$<$EXX-OEP.

\section{Theory}

In KS DFT the energy is obtained as a sum of the kinetic energy of non-interacting electrons, 
interaction with the ionic potential, the Hartree energy and the XC term 
\begin{equation}
E^{\rm KS}=T_s+E_{\rm ext}+E_{\rm H}+E_{\rm xc}\,.
\label{eq_KS_en}
\end{equation}
The KS potential for the last three terms is obtained by variation with respect to the 
electron density and serves to obtain the KS states $\varphi_i$ and single particle KS energies $\epsilon_i$
from the KS equation
\begin{equation}
(-\nabla^2/2 + v_{\rm ext} + v_{\rm H} + v_{\rm xc})|\varphi_i\rangle=\epsilon_i |\varphi_i\rangle\,.
\label{eq_ks_state}
\end{equation}
The energy in MBPT can be obtained from energy functionals such as the one of Klein\cite{klein1961} 
or the one of Luttinger and Ward.\cite{luttinger1960}
For example the energy functional of Klein evaluated at a Green's function $G$ takes the form
\begin{equation}
E^{\rm K}=E_{\rm H}[G] + i {\rm Tr}\left\{G G_{\rm H}^{-1}-1 + \ln(-G^{-1}) \right\} - i \Phi[G]\,,
\label{eq_klein}
\end{equation}
where $G_{\rm H}$ is the Hartree Green's function, ${\rm Tr}$ denotes the trace, and the so-called $\Phi$ functional
contains the perturbative electron interaction terms.\cite{barth2005,casida1995,baym1961}
The Klein functional is based on the Green's function formalism and thus appears quite different from the KS equation
on first sight.
However, as discussed in detail in Ref.~\onlinecite{barth2005}, the connection to KS DFT can be made when the functional 
is evaluated for the KS Green's function $G_0$.
Since the KS Green's function corresponds to effectively non-interacting electrons, equation~(\ref{eq_klein}) can be
much simplified, and 
 one obtains an equation identical to the KS energy formula, Eq.~(\ref{eq_KS_en}),
with the XC energy given by the $\Phi$ functional, $E_{\rm xc}=-i \Phi[G_0]$.\cite{barth2005,casida1995,baym1961}

The XC potential is obtained from the variation of the $\Phi$ functional with respect to the density $\vxc=-i\delta\Phi[G_0]/\delta \varrho$.
This can be done with chain derivatives using the basic property that $\Sigma_{\rm xc}=\delta\Phi[G_0]/\delta G_0$,
and leads to the  
Sham-Schl\"{u}ter (SS) equation,\cite{sham1983,gorling1994,casida1995,niquet2003,barth2005,gruning2006}
\begin{equation}
\begin{split}
&\int\d{\rr'}\vxc(\rr') \int {\d{\omega}\over 2\pi} G_0(\rr,\rr',\omega)G(\rr',\rr,\omega)=\\
&\int\d{\rr'}\int\d{\rr''}\int {\d{\omega}\over 2\pi} G_0(\rr,\rr',\omega)\Sigma_{\rm xc}(\rr',\rr'',\omega)G(\rr'',\rr,\omega)\,.
\label{eq_SS_eq}
\end{split}
\end{equation}
In practical implementations, the exact Green's function is replaced by the non-interacting (KS) one 
leading to the so-called linearized Sham-Schl\"{u}ter (LSS) equation.\cite{godby1988}
Then the $\omega$ integration can be performed on the left hand side of Eq.~(\ref{eq_SS_eq}) to obtain 
the independent particle response function $\chi_0(\rr,\rr')$.
The LSS equation then reads 
\begin{equation}
\begin{split}
i&\int\d{\rr'}\vxc(\rr') \chi_0(\rr',\rr)=\\
&\int\d{\rr'}\int\d{\rr''}\int {\d{\omega}\over 2\pi} G_0(\rr,\rr',\omega)\Sigma_{\rm xc}(\rr',\rr'',\omega)G_0(\rr'',\rr,\omega)\,.
\label{eq_LSS_eq}
\end{split}
\end{equation}
The SS equation can be also obtained using the property that the density
of the interacting and reference KS system are the same.\cite{sham1983}

The equation for the KS potential simplifies when only the exchange term is included in the self-energy, corresponding to the HF method.
The exchange only self-energy is frequency independent and reads
\begin{equation}
\Sigma_{\rm x}(\rr,\rr')={\sum_n^{\rm occ} \varphi_n(\rr) \varphi^*_n(\rr')\over |\rr-\rr'|}\,.
\label{eq_sigmax}
\end{equation}
When $\Sigma_{\rm x}$ is inserted into the LSS equation, the $\omega$ integration can be performed
analytically also on the right hand side leading to the EXX-OEP equation
\begin{widetext}
\begin{equation}
\int\d{\rr'}\vx(\rr') \chi_0(\rr',\rr)=
\sum_i^{\rm occ}\sum_a^{\rm unocc} \int\d{\rr'}\int\d{\rr''} \varphi_a(\rr) \varphi_a^*(\rr')  \Sigma_{\rm x}(\rr',\rr'')\varphi_i(\rr'') \varphi_i^*(\rr) 
{1\over \epsilon_i-\epsilon_a}  + c.c. \,.
\label{eq_exx_oep}
\end{equation}
\end{widetext}
The equation for the exchange potential can be also directly obtained by performing a functional derivative 
of the energy with respect to the electron density, that is $\vx(\rr)={\delta E_{\rm x}\over \delta \varrho(\rr)}$, 
as shown, e.g., in Ref.~\onlinecite{gorling1994}.

The solution of the OEP equation even in the exchange only case is still quite involved, but can be simplified by 
using the orbital shifts as suggested by K\"{u}mmel and Perdew,\cite{kummel2003,kummel2003prb}
\begin{equation}
\sum_i \psi_{i}^*(\rr) \varphi_i(\rr) + {\rm c.c.}=0\,,
\end{equation}
where $\psi_{i}$ is the linear response of the orbital $i$ to the replacement of the orbital dependent
potential with the local one.\cite{krieger1992p2,grabo2000}
The equation then simply states that the response of the density to replacing the non-local HF potential
by the local KS potential must be zero. 
 
The OEP equation can be also approximated in various ways.
Well-known approximations are the 
Krieger-Li-Iafrate approximation and the local HF method (LHF) of Della Salla and G\"{o}rling.
The first one is derived by setting the energy difference to a constant in the denominator of Eq.~(\ref{eq_exx_oep}).
This approximation gives the Slater potential $v^{\rm S}$ plus a correction term
\begin{equation}
v^{\rm KLI}(\rr)=v^{\rm S}(\rr)+{2\over \varrho(\rr)}\sum_{i\ne m}^{\rm occ} \varphi_i (\rr)\varphi_i(\rr)\langle\varphi_i| 
\vx - v^{\rm HF}| \varphi_i\rangle \,,
\end{equation}
where the highest energy occupied orbital $m$ is not included in the summation.
The Slater potential can be written as 
\begin{equation}
v^{\rm S}(\rr)={2\over \varrho(\rr)} \int \d\rr' {|\gamma(\rr,\rr')|^2 \over |\rr-\rr'|}\,,
\end{equation}
where $\gamma(\rr,\rr')=\sum_i^{\rm occ} \varphi^*_i(\rr)\varphi_i(\rr')$ is the one particle electron density matrix
for one spin.
In fact, similar approximations can be made also for the $GW$ self-energy as discussed by Casida.\cite{casida1995}
The LHF was derived by imposing that the KS and HF orbitals are identical and gives a similar
correction
\begin{equation}
v^{\rm LHF}(\rr)=v^{\rm S}(\rr)+{2\over \varrho(\rr)}\sum_{ij}^{\rm occ} \varphi_i (\rr)\varphi_j(\rr)\langle\varphi_j| 
\vx - v^{\rm HF}| \varphi_i\rangle \,.
\end{equation}
Again, only occupied states are involved in the evaluation.
Moreover, the equation has the advantage that it is invariant under unitary transformations among the occupied orbitals.

The OEP equation corresponding to the RPA total energy functional is obtained when the corresponding self-energy, 
the $GW$ self-energy, is used in Eq.~\ref{eq_LSS_eq} for $\Sigma_{\rm xc}$.
The $GW$ self-energy $\Sigma_{GW}$ is obtained from the KS Green's function and a dynamically screened Coulomb interaction $W$
\begin{equation}
\Sigma_{GW}(r,r',\omega)={i\over4\pi}\int_{-\infty}^{\infty}{\rm d}\omega' {\rm e}^{i\omega' \delta} G(r,r',\omega+\omega') W(r,r',\omega')\,,
\label{eq_GW_sigma}
\end{equation}
where $\delta$ is a positive infinitesimal.
This is similar to the HF case, however, in the HF case the bare Coulomb interaction is present and 
the frequency integration can be performed analytically leading to the self-energy given in Eq.~(\ref{eq_sigmax}).

The frequency dependence of $\Sigma_{GW}$ makes the solution of the OEP equation computationally rather involved.
To simplify the calculations, we approximate the full frequency dependency of the $GW$ self-energy by the 
self-energy operator at the poles. This method is usually referred to as self-consistent Quasi-Particle $GW$ approximation
(scQP$GW$) \cite{faleev2004,schilfgaarde2006,shishkin2007}. The matrix elements are then given by
\begin{equation}
\Sigma_{ij}^{\rm scQPGW}={1\over 2 } \langle i|  \Sigma_{GW}^*(\epsilon_i) + \Sigma_{GW}(\epsilon_j) |j \rangle \,,
\end{equation}
where the first term is complex conjugated to make the self-energy Hermitian, see Ref.~\onlinecite{shishkin2007}.
As usual, $\Sigma_{GW}(\epsilon_i)$ means that the self-energy is evaluated at the energy of the state $i$.
This leads to a considerable computational simplification of the OEP equation, since the energy dependence of the self-energy
is effectively removed.
The frequency is then present only in the Green's functions and the frequency integral can be performed analytically
in the same way as in the exchange-only case.
In fact, the approximate RPA-OEP equation is identical to the EXX-OEP equation, Eq.~(\ref{eq_exx_oep}), only the matrix elements of the 
exchange-only self-energy, $\langle i|  \Sigma_{\rm x}|j \rangle$, need to be replaced by $\Sigma_{ij}^{\rm scQPGW}$.
We represent the quantities using a plane-wave basis set so that the final equation for the RPA-OEP potential reads
\begin{equation}
\vxc(G)=\sum_{G'} \chi^{-1}_0(G,G')  \sum_{a,i} \langle a | {\rm e}^{iG'\rr} | i\rangle {\Sigma_{ia}^{\rm scQPGW} \over \epsilon_i-\epsilon_a} + c.c.\,,
\end{equation}
where $G$ and $G'$ are the wave vectors of the plane-wave basis set.
We note that without the scQP$GW$ approximation the fully frequency dependent self-energy would be required,
while in the current case only the elements around $\epsilon_i$ and $\epsilon_j$ are needed, see also Ref.~\onlinecite{shishkin2006}.

Recently Bleiziffer and coworkers derived an OEP equation for the (direct) RPA energy functional by performing 
the variation of the energy with respect to the orbitals and eigenvalues, which ultimately involves the variation 
of the response function with respect to these properties.\cite{bleiziffer2013}
This approach then avoids the MBPT route where the variation with respect to the Green's function is taken
to obtain the Sham-Schl\"{u}ter equation.
In principle, this approach and the full Sham-Schl\"{u}ter equation should be equivalent, 
but the proof of this is beyond the scope of the present work.

The fundamental or quasi-particle gap of the system is obtained as the difference of the ionization potential
and the electron affinity $E_{\rm gap}=I-A$.
These are the electron removal and addition energies, respectively.
For the orbital dependent functionals, the gap is not given by the energy difference of the eigenvalues of the single 
particle KS states but also includes a contribution of the derivative discontinuity 
$\Delta_{\rm xc}$ in the XC potential.\cite{perdew1982,perdew1983,sham1983,sham1985sch}
This is a consequence of the fact that the energy functional, unlike in the, e.g., LDA case, is not an explicit and
differentiable functional of the electron density, as discussed recently by Yang and coworkers.\cite{yang2012jcp}
The close relation between the quasi particle correction and the derivative discontinuity leads to similar 
expressions for them.
To obtain the quasi-particle energy, the quasi-particle equation
\begin{equation}
\epsilon_i^{\rm QP}=\langle i | -\nabla^2/2 + v_{\rm ext} + v_{\rm H} + \Sigma_{\rm xc}(\epsilon_i^{\rm QP})| i \rangle \,,
\label{eq_QP_energy}
\end{equation}
is linearized and after substitution from Eq.~(\ref{eq_ks_state}) one obtains the usual first order expression
\begin{equation}
\epsilon_i^{\rm QP}=\epsilon_i^{\rm KS}+Z_i\langle i | \Sigma_{\rm xc}(\epsilon_i^{\rm KS})-\vxc| i \rangle \,,
\end{equation}
where the renormalization factor $Z_i=(1-{\rm Re}\langle i| \frac{\partial}{\partial \omega} \Sigma(\omega)\big|_{\epsilon_i^{\rm KS}} | i \rangle)^{-1}$
stems from the fact that the self-energy should be evaluated at the quasi particle energy and not at the KS energy.
For the derivative discontinuity, Niquet and Gonze\cite{niquet2004} argued that the self-energy should be taken at the KS single 
particle energy so that there is no renormalization factor 
\begin{equation}
\epsilon_i^{\rm \Delta}=\epsilon_i^{\rm KS}+\langle i | \Sigma_{\rm xc}(\epsilon_i)-\vxc| i \rangle \,.
\end{equation}
The final correction to the gap is then obtained by subtracting the corrections
to the two states between which  the gap is calculated
\begin{equation}
\Delta_{\rm xc}=\epsilon_i^{\rm KS}+\langle i | \Sigma_{\rm xc}(\epsilon_i)-\vxc| i \rangle 
- \epsilon_j^{\rm KS}-\langle j | \Sigma_{\rm xc}(\epsilon_j)-\vxc| j \rangle\,,
\end{equation}
typically the orbital index $i$ corresponds to the conduction band minimum and $j$ to the valence band maximum.

The subtle difference in the equations for the QP gap and the true KS gap is caused by the 
use of the KS non-interacting Green's function, when the equation 
for the derivative discontinuity is derived, see, e.g., Ref.~\onlinecite{hellgren2012pra}.
More specifically, the unoccupied state to which the electron is added is assumed to have the energy of the KS
unoccupied state.
If the exact Green's function were used instead, the electron would be added with the quasi-particle energy, 
that is, with the energy given by Eq.~(\ref{eq_QP_energy}).
Then the renormalization factor would be present in the formula for the derivative discontinuity in the XC potential
and the true KS gap would be identical to the quasi-particle gap calculated from MBPT. This suggests that
the renormalization factor should be included.
The use of the renormalization factor to obtain $\Delta_{\rm xc}$ is also supported by calculations in Ref.~\onlinecite{lannoo1985}, 
where the derivative discontinuity was obtained for a two plane-wave model from the exact Green's function for that model system.
Indeed, Eqs. (48) and (49) in Ref.~\onlinecite{lannoo1985} are essentially the same as the standard equations for 
the gap obtained from the MBPT approach, that is, including the renormalization factor, which is the denominator of Eq. (49) in Ref.~\onlinecite{lannoo1985}.
We will, nevertheless, report the gaps calculated using both $\epsilon_i^{\rm QP}$ and $\epsilon_i^{\rm \Delta}$ in this work.

\section{Computational setup}

The calculations have been performed using the plane-wave code VASP.\cite{kresse1996} 
The OEP routines use parts of the Hartree-Fock and $GW$ routines which were described 
previously.\cite{paier2005,paier2006,gajdos2006,shishkin2006}
The one center terms of the projector-augmented-wave (PAW) method\cite{blochl1994,kresse1999} are obtained using LHF
and kept fixed for subsequent EXX-OEP and RPA-OEP calculations.

It is well-established that the properties obtained from MBPT, such as the band gap correction, depend more strongly on the
technical parameters than the ground state KS calculations.\cite{schilfgaarde2006}
For example, the plane-wave basis set cut-off $E_{\rm cut}^{\rm PW}$, the cut-off employed for the response function $E_{\rm cut}^{\rm \chi}$,
the number of bands included in the calculation of the response function, and the number of k-points are important parameters.
For the OEP calculations we additionally need to control the convergence with the size of the self-energy 
matrix which is represented in a basis set of KS states.

\begin{figure}[h]
\centerline{\includegraphics[height=5.cm]{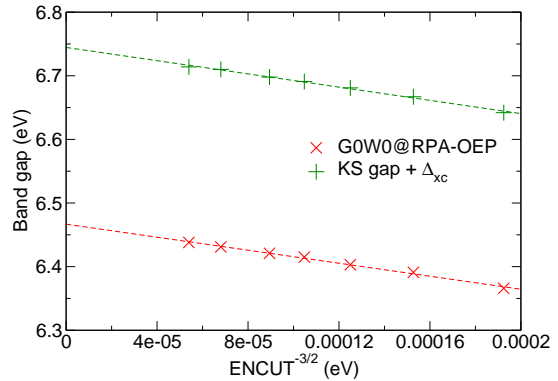}}
\caption{Convergence of the G$_0$W$_0$@RPA-OEP band gap (red $\times$) and of the KS gap with the derivative discontinuity
correction (green $+$) for diamond $\Gamma$$\rightarrow$X transition. 
The data are plotted against the energy cut-off for the response function to the power of $-3/2$ which
corresponds to the inverse of the number of basis functions used in the response function.
The data are fitted with a linear function which closely follows the data.}
\label{fig_C_convergence}
\end{figure}

The number of unoccupied bands $N^{\rm bands}$ used to obtain the response function is an important parameter 
in the convergence of the band gaps. 
The error decreases as $1/N^{\rm bands}$ and to reduce the errors we use all 
the unoccupied bands spanned by the basis set for a given $E_{\rm cut}^{\rm PW}$.\cite{shepherd2012}
Furthermore, the response function and the screened interaction $W$ are stored using a plane-wave basis
with an energy cut-off $E_{\rm cut}^{\rm \chi}$.
The number of basis set functions $N^{\rm \chi}$ is then proportional to $(E_{\rm cut}^{\rm \chi})^{(3/2)}$
and the error in the band gap corrections also decreases like $1/N^{\rm \chi}$, similar to the convergence
of the total energy observed in, e.g., Ref.~\onlinecite{harl2008}.
To obtain converged results we perform a set of calculations for several values of $E_{\rm cut}^{\rm PW}$ 
and consistently set $E_{\rm cut}^{\rm \chi}=2/3 E_{\rm cut}^{\rm PW}$.
Finally, we perform an extrapolation to the infinite basis set limit by assuming that the error decreases as
 $(E_{\rm cut}^{\rm \chi})^{-(3/2)}$.
Overall, our procedure is similar to the one used in Ref.~\onlinecite{garcia2007}.
We also note that a similar relation between $E_{\rm cut}^{\rm \chi}$ and $E_{\rm cut}^{\rm PW}$ was used 
in Ref.~\onlinecite{stadele1999}, the basis set extrapolation was, however, not performed in that study.
An example of the convergence for the $\Gamma$$\rightarrow$X gap in diamond is shown in Figure~\ref{fig_C_convergence} together
with a linear fit to the data.

The data also depend on the number of k-points used;
we observed that increasing the number of k-points generally reduces the QP gap and increases the OEP gap.
For example, the $\Gamma$$\rightarrow$X $G_0W_0$@RPA-OEP gap in Si changes from 1.28~eV to 1.35~eV upon increasing
the k-point sampling from $4\times4\times4$ to $8\times8\times8$, when the same KS potential is used.
This convergence behavior is caused by the treatment of the divergence of the Coulomb potential at the $\Gamma$ point.
While we properly deal with  the divergence in the potential, there are non-divergent regular terms that are not accounted for,
leading to a residual error.
This error is proportional to the volume per k-point in the reciprocal space and,
therefore, decreases as $1/N_{\rm kpt}^3$, where $N_{\rm kpt}$ is the number of k-points in one direction,
see, e.g. Ref~\onlinecite{schilfgaarde2006}.
To obtain results converged with respect to the k-point sampling we extrapolate the results obtained 
for the $4\times4\times4$ and $6\times6\times6$ k-point grids.
This scheme is accurate as observed by a calculation with an  $8\times8\times8$ k-point grid for diamond.
In many materials the convergence with the basis set cut-off shows an opposite trend, 
both the QP and KS gaps increase with increasing basis set.\cite{note_gaps_increase} 
Therefore the usually employed small basis and k-point grids can lead to similar results for the QP gaps
as the data converged with respect to both k-point sampling and basis-set completeness.

Another parameter that affects the results is the number of KS states that are used to represent 
the self-energy matrix in the OEP equation.
While using more bands gives a higher accuracy, it also increases the computational time.
We therefore set the number of states such that an accuracy of about 0.01~eV in the QP band gap is obtained.
We typically use 80 to 112 KS states.
The number of iterations of the self-consistent OEP procedure was set to 
10 which also guarantees convergence to better than 0.01 eV.
Overall, we expect our values to be converged typically well within 0.05~eV.

\section{Materials and PAW potentials}

For the present study we considered a range of materials, semiconductors, noble gas solids, and ionic solids.
The materials were studied at their experimental lattice parameter and 
the cubic diamond or zinc blende structure was used for the semiconductors.
The PAW potentials used for the different materials are given in Table~\ref{tab_paws},
these are generally the ``GW" type PAW potentials that include projectors at energies above the vacuum level.
For some materials we tested more potentials and then more entries are shown.
The spin-orbit coupling was not included in the present calculations.

\begin{table}[h]
\caption{The PAW potentials used in this study using their names in the VASP distribution.
For some of the materials we obtained the results using two different potentials for comparison.
Second column gives the number of valence electrons, third column
lists the number of partial waves and projectors for a given angular momentum (s, p, d, f). 
The local potential (fourth column) is either a potential obtained for a high angular momentum (d, f)
or a truncated, pseudized all electron potential.
In the latter case the cut-off radius for the local potential in a.u. is given in the table.
The default plane-wave basis cut-off is shown in the last column.
}
\label{tab_paws}
\centerline{
\begin{ruledtabular}
\begin{tabular}{llcccc}
& PAW potential & N. val. el. & Projectors & Local & $E_{\rm cut}$ (eV) \\
\hline
& Al\_sv\_GW & 11& 3s 3p 2d &f & 411\\
&Al\_GW & 3& 2s 2p 2d &f & 241 \\
& Ar\_GW & 8 & 2s 2p 2d &f & 266\\
& As\_GW& 5 &2s 2p 1d &f & 209\\
&As\_d\_GW & 15& 2s 2p 2d & f& 864\\
& B & 3& 2s 2p & d& 319\\
& C\_GW & 4 & 2s 2p & d& 414 \\
& F\_GW\_new & 7 & 3s 3p 2d & f& 487\\
& Ga\_d\_GW & 13& 2s 2p 3d & f& 405\\
& Ge\_d\_GW & 14& 2s 2p 2d &f & 310\\
& Ge\_sv\_GW & 22 & 3s 3p 2d &f & 455\\
& Li\_sv\_GW& 3 & 3s 2p 1d & 1.003 & 433\\ 
& Mg\_sv\_GW& 10 & 3s 3p 2d &1.204 & 430 \\ 
& N\_GW& 5 & 2s 2p & d& 421\\
& Ne\_GW & 8& 2s 2p &d & 318 \\
& O\_GW&  6& 2s 2p & d& 414 \\
& O\_GW\_new& 6 & 3s 3p 2d & f& 434 \\
& P\_GW& 5 & 2s 2p 2d  &f & 255\\
& Si\_d\_GW& 4 & 2s 2p 2d & f& 246 \\
& Si\_sv\_GW& 12 & 3s 3p 2d 1f & 1.712 & 547\\ 
& Zn\_sv\_GW(1f) & 20 & 3s 3p 3d 1f&f & 499\\
& Zn\_sv\_GW(2f)& 20& 3s 3p 3d 2f&1.206 & 496 \\
\end{tabular}
\end{ruledtabular}}
\end{table}

\section{Results}

\subsection{Band gaps}
\label{sec_gaps}

\begin{figure}[h]
\centerline{\includegraphics[height=5.5cm]{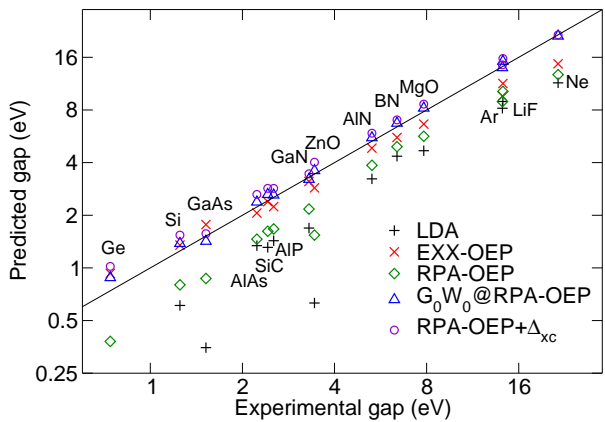}}
\caption{The single particle and quasi-particle gaps for different materials compared to experimental reference data.
The Kohn-Sham single particle gaps are shown for the LDA, EXX-OEP, and RPA-OEP methods. 
The quasi-particle values based on RPA-OEP are shown both with the $Z$ renormalization factor, corresponding to $G_0W_0$@RPA-OEP, 
and without $Z$. (Note the double logarithmic scale.)} 
\label{fig_gaps}
\end{figure}

We now discuss our results starting with the Kohn-Sham and quasi-particle gaps as obtained with 
different methods. 
The values extrapolated to an infinite number of k-points and complete basis set are shown in Tables~\ref{tab_results} and~\ref{tab_results2}.
We compare to experimental data and previous EXX-OEP calculations, where available.
The minimal gaps are also compared in Figure~\ref{fig_gaps}.

For the KS EXX-OEP gaps we can compare to the data obtained with the all-electron full-potential linearized 
augmented-plane-wave (FLAPW) method by Betzinger~{\it et al.} in Refs.~\onlinecite{betzinger2011} 
and~\onlinecite{betzinger2012}.
We generally find good agreement with these results, for example, the difference for the gaps of noble gas solids, 
Ge, or Si is around 0.05~eV.
In the case of BN, AlN, and SiC there is a larger difference for some of the gaps, up to $\sim$0.3~eV,
the smallest energy gap, however, differs in all these cases by less than $\sim$0.1~eV.
While our results are generally in reasonable agreement with previously published pseudopotential
calculations, a direct comparison is often not possible, since the published values usually include LDA correlation,
that we have not included here since our goal is to include RPA correlation.

It turns out that the results within the current implementation are in some cases somewhat
sensitive to the number of valence electrons used in the PAW potential. 
For example, the $\Gamma$$\rightarrow$L gap of Si increases by about 0.18~eV when the 2$s$ and 2$p$
shells are included in the valence. 
Interestingly, the effect of increasing the number of valence electrons for Al is much smaller;
we observe the largest change in the gap, 0.05~eV, for the $\Gamma$$\rightarrow$$\Gamma$ transition of AlAs
when the number of valence electrons is increased from 3 to 11.
The elements with $d$ electrons turn out to be the most problematic cases in this regard.
For GaAs, we found it necessary to include the 3$d$ electrons of As to obtain the direct gap as the 
lowest transition.
For Ge, the 3$s$, 3$p$, and 3$d$ shells need to be included to obtain the same energy ordering of the gaps 
as in the FLAPW calculations. 
Since this Ge potential possesses 22 valence electrons and we had to use cut-offs between 350~eV and 500~eV for the 
response function, the calculations are computationally rather demanding even though the cell 
contains only two atoms.

\begin{table*}[t]
\begin{minipage}{\textwidth}
\caption{Kohn-Sham and quasi-particle transition energies between the $\Gamma$-point and the indicated k-point  
compared to previous results (column ``EXX other") and experimental reference where available. 
The ``Z" scaling factor for the self-energy correction is included in the $G_0W_0$ column,
while unscaled data are shown in the ``OEP$+\Delta_{\rm xc}$" column for the RPA-OEP method.
The PAW potential used to obtain the data is given in the PAW column and all the results have been
extrapolated with respect to the number of k-points and plane-wave basis set cut-off. All data in eV.}
\footnotesize
\label{tab_results}
\centerline{
\begin{ruledtabular}
\begin{tabular}{lllcccccccc}
Solid     & k-point& PAW & LDA & \multicolumn{3}{c}{EXX} &  \multicolumn{3}{c}{RPA} & Expt. \\
          &        &    &     &   OEP    &OEP other  &  OEP$+\Delta_{\rm x}$   &   OEP    &  $G_0W_0$  &   OEP$+\Delta_{\rm xc}$      & \\
\hline
C & $\Gamma$   &  C\_GW &5.54  & 6.20 & 6.21\footnotemark[1]&14.27 &6.07 & 7.68 & 8.00& 7.3\footnotemark[3]\\
  & L        &         & 8.38& 9.07&9.09\footnotemark[1] &17.99  & 8.92  & 10.64 & 10.99&\\
  & X        &         & 4.71& 5.36&5.20\footnotemark[1] &12.90  & 5.00 & 6.43  & 6.70&\\
Si& $\Gamma$   & Si\_d\_GW &2.53 &3.13& 3.13\footnotemark[1]&8.52& 2.69  & 3.32 & 3.52  &  3.05\footnotemark[4]\\
  & L        &           &1.43 &  2.11& 2.21\footnotemark[1]& 7.16  &  1.60 & 2.23 &  2.42  & 2.4\footnotemark[5]\\
  & X        &           & 0.61 &  1.26 & 1.30\footnotemark[1]& 6.02  & 0.73 &1.37  & 1.56 & 1.25\footnotemark[4]\\
Si& $\Gamma$   & Si\_sv\_GW & 2.53 &3.15  &3.13\footnotemark[1] & 8.53  &2.69  &3.33 & 3.54&3.05\\
  & L        &          &1.43  &2.29 & 2.21\footnotemark[1]&  7.28 & 1.75 & 2.25&2.39 &2.4\\
  & X        &          &0.61 & 1.35  & 1.30\footnotemark[1]&  6.04 &0.80 &1.38 &1.54 &1.25\\
SiC&$\Gamma$  &Si\_d\_GW &6.35  &7.50 &7.18\footnotemark[1]&14.97  & 6.83   &  7.61& 7.81 &    \\
   &L        & C\_GW    & 5.44 &6.30 &6.14\footnotemark[1]&13.60  &5.83 & 6.88 &  7.14 &\\
   &X        &          & 1.31& 2.39  &2.29\footnotemark[1]&8.39 & 1.62 & 2.63 & 2.85 &2.42\footnotemark[6]\\
BN &$\Gamma$   & B\_GW    &8.69 &  9.84 &9.80\footnotemark[2]&19.14&  9.51 & 11.69  & 12.10 &     \\
   &L        & N\_GW    & 10.21& 11.15 &10.88\footnotemark[2]&20.50& 10.76 & 12.56 &  12.92 &\\
   &X        &          & 4.35& 5.57 &5.42\footnotemark[2]&13.48 & 4.93 &6.72 & 7.02 & 6.4\footnotemark[7]\\
AlN&$\Gamma$ & Al\_sv\_GW    &4.22 &  5.69 &5.46\footnotemark[2]&13.27 & 4.85 & 6.31  & 6.65  & 5.93\footnotemark[8]\\
   &L       & N\_GW       &7.25 &8.59 &8.42\footnotemark[2]&16.97&  7.94 & 9.76 & 10.16&\\
   &X       &            & 3.23&4.84 &4.77\footnotemark[2]&11.90 & 3.86 &  5.55 &  5.89 &  5.3\footnotemark[8]\\
GaN &$\Gamma$ &  Ga\_d\_GW  & 1.69& 3.12 &3.11\footnotemark[2]& 9.86 & 2.17 & 3.21 & 3.45 &3.30\footnotemark[9]\\
   &L       &  N\_GW    & 4.48&  5.83 &5.94\footnotemark[2]&13.51 &  5.05  & 6.39 & 6.68& \\
   &X       &           & 3.25& 4.67 &4.61\footnotemark[2]&11.25 & 3.70 & 4.96& 5.21  &  \\
AlP &$\Gamma$& Al\_GW   & 3.16 & 4.29 &&10.09&  3.66 & 4.47 & 4.70 &  \\
   &L       & P\_GW    & 2.69 &  3.52 && 9.27&  3.08 &  4.00 & 4.26  & \\
   &X       &          & 1.43&  2.24  &&7.50 & 1.67 & 2.61 & 2.85&2.53\footnotemark[10]\\
AlP &$\Gamma$& Al\_sv\_GW   & 3.16 & 4.27 & & 10.13   & 3.57   &  4.44  &   4.69    &  \\
   &L       & P\_GW    & 2.69 &       3.51 & &  9.32   & 3.01  & 3.98  & 4.25   & \\
   &X       &          & 1.43&     2.21   & &   7.49  &   1.68 &2.64   & 2.89 &2.53\footnotemark[10]\\
AlAs&$\Gamma$&  Al\_GW  &1.88  &3.28 &3.20\footnotemark[11] &  8.40& 2.47 &3.20 &3.39 &3.13\footnotemark[5]\\
   &L       &  As\_GW    & 2.02 &2.92 &2.99\footnotemark[11] &  8.17 & 2.38 &3.18 &3.40 &2.57\footnotemark[5]\\
   &X       &            & 1.34 & 2.10 &2.26\footnotemark[11] &  7.08 & 1.48 &2.31 &2.55 &2.23\footnotemark[5]\\
AlAs&$\Gamma$&  Al\_sv\_GW  &1.88  &3.33 &3.20\footnotemark[11] & 8.47 &2.57 &3.24 & 3.43&3.13\footnotemark[5]\\
   &L       &  As\_GW     & 2.02 &2.90 &2.99\footnotemark[11] & 8.26  &2.31 &3.20 &3.45 &2.57\footnotemark[5]\\
   &X       &            & 1.34 &2.06  &2.26\footnotemark[11] &7.10   &1.46 &2.38 &2.63 &2.23\footnotemark[5]\\
\end{tabular}
\end{ruledtabular}
}
\begin{minipage}[t]{0.49\linewidth}
\footnotetext[1]{Ref.~\onlinecite{betzinger2011}}
\footnotetext[2]{Ref.~\onlinecite{betzinger2012}}  
\footnotetext[3]{Ref.~\onlinecite{roberts1967}}  
\footnotetext[4]{Ref.~\onlinecite{ortega1993}} 
\footnotetext[5]{Ref.~\onlinecite{lb_data_semic}}  
\footnotetext[6]{Ref.~\onlinecite{humpreys1981}}  
\end{minipage}\hfill
\begin{minipage}[t]{0.49\linewidth}
\footnotetext[7]{Ref.~\onlinecite{chrenko1974}} 
\footnotetext[8]{Ref.~\onlinecite{roppischer2009}} 
\footnotetext[9]{Refs.~\onlinecite{okumura1997} and~\onlinecite{ploog1997}}
\footnotetext[10]{Ref.~\onlinecite{lorenz1970}}
\footnotetext[11]{Ref.~\onlinecite{aulbur2000}}
\end{minipage}
\end{minipage}
\end{table*}

\begin{table*}[t]
\begin{minipage}{\textwidth}
\caption{Same as Table~\ref{tab_results}.}
\footnotesize
\label{tab_results2}
\centerline{
\begin{ruledtabular}
\begin{tabular}{lllcccccccc}
Solid     & K-point& PP & LDA & \multicolumn{3}{c}{EXX} &  \multicolumn{3}{c}{RPA} & Expt. \\
          &        &    &    &   OEP    &OEP other  &  OEP$+\Delta_{\rm x}$  &   OEP    &  $G_0W_0$  &  OEP$+\Delta_{\rm xc}$      & \\
\hline
Ge &$\Gamma$  & Ge\_d\_GW  & -0.14 &  1.14 &1.24\footnotemark[1]& 5.41 & 0.31 & 0.62 &  0.72 & 0.9\footnotemark[2]\\
   &L       &           & 0.07 & 0.88  &0.89\footnotemark[1]&5.00 & 0.29 & 0.71 & 0.82 &  0.74\footnotemark[2]\\
   &X       &           & 0.67 & 1.20 &1.15\footnotemark[1]& 5.45 & 0.73 &1.23& 1.37 &  1.3\footnotemark[3]\\
Ge &$\Gamma$  & Ge\_sv\_GW  &-0.15  & 1.28      &1.24\footnotemark[1]& 5.76 &0.50    &0.98   &1.11   &  0.9\footnotemark[2]\\
   &L       &           & 0.06& 0.93     &0.89\footnotemark[1]& 5.19  &  0.38   & 0.88   &  1.02   &  0.74\footnotemark[2]\\
   &X       &           & 0.67 &  1.20   &1.24\footnotemark[1]& 5.51   &  0.72    & 1.28   & 1.44   & 1.3\footnotemark[3]\\
GaAs &$\Gamma$  & Ga\_d\_GW  &0.35 &  1.77& 1.72\footnotemark[1] & 6.55 & 0.87 &1.42 &1.57  &1.52\footnotemark[4] \\ 
   &L       &   As\_d\_GW  &0.89  &  1.88 &1.79\footnotemark[1] & 6.65 & 1.15 &1.81 & 1.99    &      \\
   &X       &           &1.36 &    2.07  & 1.95\footnotemark[1] & 6.71 & 1.44 &2.16 & 2.35   & \\ 
Ne &$\Gamma$  & Ne\_GW   & 11.43&14.69 &14.79\footnotemark[1]&25.65& 12.73 & 21.16 &21.49& 21.51\footnotemark[5]\\
   &L       &          &17.13 & 20.45 &20.49\footnotemark[1]&32.04&  18.56 &27.13 & 27.56&\\
   &X       &          & 18.32& 21.80 &21.85\footnotemark[1]&32.60&  19.78 & 28.01 & 28.39&\\
Ar &$\Gamma$  & Ar\_GW  & 8.18&9.61 &9.65\footnotemark[1]&18.12  &8.93 &   13.96&  14.48  &  14.15\footnotemark[5]\\
   &L       &         & 11.05&  12.18 &12.22\footnotemark[1]&21.46&  11.67& 17.13&17.72&\\
   &X       &         & 10.85&  12.05 &12.08\footnotemark[1]&21.05&  11.55  &  16.91  & 17.45&\\
LiF &$\Gamma$ &  Li\_sv\_GW& 8.94& 11.33 &&22.04 &10.21 &15.19&15.75&14.20\footnotemark[6]\\
   &L       &  F\_GW\_new    & 10.45 & 13.11 &&23.77  &11.86&16.92&17.51&\\
   &X       &          & 15.54 & 17.02 &&28.89   &  15.88&21.25&  22.08&\\
MgO &$\Gamma$ &  Mg\_sv\_GW  &4.68  &6.64 &&15.46 & 5.66   &  8.20&  8.64&7.83\footnotemark[7]\\
   &L       &  O\_GW     & 7.76 & 9.70 &&18.84  &8.71&11.52 &12.00&\\
   &X       &          & 8.91 & 10.89 &&19.79   &   9.76 &   12.70  &   13.16&\\
ZnO &$\Gamma$  & Zn\_sv\_GW(1f)&0.63 & 2.83 &&11.09  & 1.39& 3.54& 3.98& 3.44\footnotemark[8]\\
   &L       &    O\_GW        & 5.33            &7.34 &&17.03   &6.05 & 8.75& 9.30&\\
   &X       &                  & 5.13         & 7.52 && 15.64  & 6.01&8.15 &8.56 &\\
ZnO &$\Gamma$  & Zn\_sv\_GW(2f)  & 0.63 &  2.87  &&11.05 &1.54 & 3.60& 4.02&3.44\footnotemark[8]\\
   &L       &    O\_GW\_new    & 5.33 &    7.40   && 17.00 &6.20 &8.77 &9.29 & \\
   &X       &                   & 5.13 &    7.49   && 15.56 &6.09 &8.19 &8.59 &  \\
\end{tabular}
\end{ruledtabular}}
\begin{minipage}[t]{0.49\linewidth}
\footnotetext[1]{Ref.~\onlinecite{betzinger2011}}
\footnotetext[2]{Ref.~\onlinecite{wagner1984}}
\footnotetext[3]{Ref.~\onlinecite{lb_data_semic}}  
\footnotetext[4]{Ref.~\onlinecite{shishkin2007}} 
\end{minipage}\hfill
\begin{minipage}[t]{0.49\linewidth}
\footnotetext[5]{Ref.~\onlinecite{runne1995}}
\footnotetext[6]{Ref.~\onlinecite{piacentini1976}}
\footnotetext[7]{Ref.~\onlinecite{whited1973}}
\footnotetext[8]{Data for wurtzite structure.}
\end{minipage}
\end{minipage}
\end{table*}

When we compare the band gaps obtained within the KS scheme, i.e., LDA, EXX-OEP, and RPA-OEP,
we find that for every material LDA predicts the smallest gap
followed by RPA-OEP and EXX-OEP, which predicts the largest gaps. 
This is true for all the data given in Tables~\ref{tab_results} and~\ref{tab_results2} and can be also seen in Figure~\ref{fig_gaps}.
While LDA  underestimates the gap considerably, EXX-OEP data are quite close to the experimental reference
for gaps below $\approx$3.5~eV, as observed before.\cite{stadele1997,stadele1999,aulbur2000}
However, this agreement can be only accidental, and as the band gap increases
the agreement becomes worse.\cite{magyar2004}
For example, EXX-OEP predicts a gap of 14.69~eV for solid Ne compared to the experimental value of 21.51~eV.
Moreover, there is no physical reason why the EXX-OEP gaps should be similar to the experimental quasi-particle gaps.
The agreement must then be caused by an accidental cancellation between the correlation contribution to the OEP 
potential and the derivative discontinuity.
The true EXX gaps that include the derivative discontinuity correspond to the HF energy levels 
obtained with the EXX orbitals and are close to the HF gaps, much larger than the EXX-OEP or experimental gaps.
We list the true EXX gaps in the ``OEP$+\Delta_{\rm x}$" column of Tables~\ref{tab_results} and~\ref{tab_results2}.
The smaller RPA-OEP gaps compared to the EXX-OEP values can be attributed to a smaller difference between the potential
in the bond and interstitial regions and we return to this point in more detail later.
We note that Ge, which is incorrectly described to be metallic within LDA, becomes a semi-conductor in RPA-OEP or EXX-OEP.
Betzinger~{\it et al.}\cite{betzinger2012} observed similar improvements for ScN and InN when going from LDA to EXX-OEP
and thus the EXX-OEP and RPA-OEP schemes are likely to give a qualitatively better description also for other materials 
incorrectly predicted to be metallic within LDA.

The property that should be compared to the experimental data are the quasi-particle gaps 
(column ``$G_0W_0$" in Tables~\ref{tab_results} and~\ref{tab_results2}).
One can see from Tables~\ref{tab_results} and~\ref{tab_results2} and Figure~\ref{fig_gaps} that the $G_0W_0$@RPA-OEP scheme predicts 
gaps in relatively good agreement with the experimental reference.
Overall for semiconductors, a tendency to overestimate the gap is observed with 
the exception of the materials containing Ga (GaN and GaAs) where the predicted gap is too small.
For GaN there is also another experimental value available for the band gap,\cite{lei1992} which is 0.1~eV lower than
the data we compare to, and here the agreement with our calculations would be improved.\cite{okumura1997,ploog1997}
The errors are quite similar in the other cases, for example, for AlAs, SiC, and AlP, materials with a minimum
gap of about 2.4~eV the error in the $\Gamma$$\rightarrow$X gap is $\sim$0.2~eV.
We observe $\sim$5~\% too large gaps also for the ionic solids MgO and LiF while the gap is about 1--2~\% too low
for the noble gas solids Ne and Ar. 
The KS gaps including the derivative discontinuity $\Delta_{\rm xc}$ (column OEP$+\Delta_{\rm xc}$ in Tables~\ref{tab_results} and~\ref{tab_results2}),
that are obtained when the renormalization 
factor is not used, increase the predicted gaps, usually by several tenths of an eV compared to the $G_0W_0$@RPA-OEP data.

We now compare our RPA-OEP calculations to the data previously published, specifically to Ref.~\onlinecite{gruning2006}.
While we find good agreement for Ar and Si, our data are larger by at least 0.5~eV for LiF.
This difference is larger than the differences between the EXX-OEP data, which agree to about 0.1~eV.
It is difficult to pin down exactly the origin for the large difference for LiF, we note, however, 
that the plasmon-pole approximation for the dielectric function was employed in Ref.~\onlinecite{gruning2006}. 
Moreover, the data in Ref.~\onlinecite{gruning2006} are not fully converged with respect to the k-point
sampling and basis set size.

It is interesting to determine how the $G_0W_0$ results change when the RPA-OEP single particle orbitals and 
energies are used as compared to the standard input based on local or semi-local functionals.
Comparing to data previously published, we find that using RPA-OEP as an input leads to larger values 
for the gaps, and tests for Si, C, and BN show that we obtain values similar to self-consistent $GW$ data,
where the single particle energies are updated in both the Green's function and in the dielectric matrix, 
and which is known to overestimate the gaps.\cite{shishkin2007}
Actually, the standard $G_0W_0$ calculations based on semi-local functionals exhibit two properties:
i) the band gaps tend to be underestimated in general and ii) the errors in the gaps are not constant but 
depend on the  quality of the DFT functional in describing the properties of the materials.
For example, it was observed that the errors in the gap approximately follow the errors in the dielectric constant $\epsilon$.
When the gap is more strongly underestimated, the screening (and $\epsilon$) is overestimated and the QP
corrections are too low.\cite{shishkin2007}
Our calculations show that RPA-OEP does not suffer from the second point and the errors in $\epsilon$
are more consistent between materials.
Indeed, as shown in Table~\ref{tab_epsilon}, the dielectric constants are consistently underestimated.
The screening is then too weak and the $G_0W_0$@RPA-OEP gaps tend to be too large.
While we have obtained a consistent starting point for the $G_0W_0$ calculations, there is little 
benefit in the final absolute accuracy; one would have to go beyond the $GW$ approximation to improve the agreement 
with the experiment.

\begin{table}[h]
\caption{The RPA dielectric constants of solids as obtained with the RPA-OEP procedure compared to the 
experimental values.
A 6$\times$6$\times$6 k-point sampling was used to obtain the data.
}
\label{tab_epsilon}
\centerline{
\begin{ruledtabular}
\begin{tabular}{lcc}
Material     & $\epsilon^{\rm RPA-OEP}$ & $\epsilon^{\rm expt.}$ \\
\hline
C   & 5.25 & 5.70 \\
BN  & 4.11 & 4.50 \\
SiC & 6.07 & 6.52 \\
Si & 10.87 & 11.90 \\
AlP & 6.99 & 7.54 \\
GaN & 5.03 & 5.30 \\
Ge & 14.67  & 16.2 \\
ZnO & 3.67 & 3.74 \\
LiF & 1.87 & 1.90 \\
MgO & 2.77 & 3.00 \\
\end{tabular}
\end{ruledtabular}}
\end{table}

Before closing this section, we discuss the case of ZnO which is problematic since the LDA gap is
very small and the convergence of the quasi-particle gap is very slow with the size of the basis set 
used.\cite{shishkin2007,shih2010,friedrich2011}
For the zinc-blende structure our calculations give an LDA gap of only 0.63~eV, while it increases to 1.54~eV for RPA-OEP and 
again considerably to 2.87~eV using the EXX-OEP method.
This increase in the KS gap leads to a better agreement of the dielectric constant with the experimental
data compared to the RPA value of  $\epsilon^{\rm RPA-PBE}=5.12$  based on the Perdew-Burke-Ernzerhof XC functional\cite{perdew1996}
 presented in Ref.~\onlinecite{shishkin2007}.
This improves the value of the $G_0W_0$ gap which, after k-point and basis set extrapolations, is 3.60~eV.
Therefore, even for ZnO, the final error in the gap is similar to the errors obtained for other semiconductors
or ionic solids.
We note that performing only k-point extrapolation (or using, say an 8$\times$8$\times$8 k-point grid) 
and no basis-set extrapolation, as is the usual practice, would give a value at least 0.15~eV lower.
For ZnO the increase of the gap relates to a better description of the 3$d$ shell of Zn.
The 3$d$ electrons are moved to more negative binding energies in the OEP procedure, which reduces the interaction
between the O $p$ and Zn $d$ electrons and concomitantly increases the gap between the O $p$ dominated valence band states
and Zn $s$ conduction band states.
Although not quite as pronounced as for ZnO, GaAs shows a similar behavior with a significant increase of the KS
gap from LDA to RPA-OEP.

\subsection{Valence band widths}

While the EXX-OEP band gaps are consistently larger than RPA-OEP and LDA, a different situation occurs for
the valence band widths (VBW). 
As shown in Table~\ref{tab_vbw_comp}, the VBW as obtained with EXX-OEP are generally 
narrower than the RPA-OEP band widths,
for example, the difference amounts to 0.21, 0.28, and 0.25 eV for C, Si, and Ge, respectively.
In the case of Ne and Ar, the two methods give VBWs which are almost identical (to within 30 meV).
This is expected since screening in Ne and Ar is very weak and thus the EXX-OEP potential
is similar to the RPA-OEP potential.
For semiconductors the LDA and RPA-OEP VBWs are rather similar and the EXX-OEP are smaller.
Taking the VBW of Ge as an example, the values obtained with LDA, RPA-OEP, and EXX-OEP are 
12.78, 12.69, and 12.44 eV respectively. 
The reduced VBWs in the EXX-OEP case have been attributed to a more localized electron 
density.\cite{aulbur2000,stadele1999}
The localization is reduced in the RPA-OEP case, and the VBWs increase back to a value similar
to the LDA data.
The k-point mesh has a small effect on the KS values with the change in the EXX-OEP VBW 
being larger (the change is 0.02~eV for EXX-OEP and 0.005~eV for RPA-OEP for C when going from 
4$\times$4$\times$4 to 6$\times$6$\times$6 k-point sampling).

\begin{table}[h]
\caption{Valence band widths in eV for different materials as obtained with the LDA, EXX-OEP, and RPA-OEP method.
The values depend very weakly on the cut-off and k-point sampling employed, hence we show the data obtained
for 6$\times$6$\times$6 k-points.
}
\label{tab_vbw_comp}
\centerline{
\begin{ruledtabular}
\begin{tabular}{lcccc}
Material     & LDA & RPA-OEP & EXX-OEP &  \\
\hline
C   &  21.31& 21.63  & 21.42 & \\    
Si  & 11.96 & 11.90  & 11.66 &  \\ %
Ge  & 12.81 & 12.70  & 12.44 &  \\ %
GaAs& 12.79 & 12.70  & 12.36 &  \\  %
SiC & 15.36 & 15.47  & 15.25 & \\ %
Ne & 22.90 &  23.28 & 23.30 & \\ %
Ar & 14.63 & 14.95& 14.92  & \\ %
\end{tabular}
\end{ruledtabular}}
\end{table}

In Table~\ref{tab_vbw} we compare our EXX-OEP band-widths at $\Gamma$ with the results published 
previously for some materials.\cite{stadele1999,aulbur2000,fleszar2001}
Generally we find good agreement with deviations on the order of several tens of meV.
We note that in some cases the band-width is somewhat dependent on the PAW potential employed.
For example, the quoted value of 11.66~eV for Si has been obtained with the Si potential with 12 valence
electrons as given in Table~\ref{tab_paws}.
Taking only the 3$s$ and 3$p$ orbitals as valence, the band-width increases to 11.77~eV.
The value with 12 valence electrons is, in fact, ``technically" more accurate and agrees well
with previous calculations.

\begin{table}[h]
\caption{EXX-OEP valence band widths in eV at $\Gamma$ for different materials compared to previously published results.
}
\label{tab_vbw}
\centerline{
\begin{ruledtabular}
\begin{tabular}{lcccc}
Material     & This work & Ref.~\onlinecite{stadele1999}&Ref.~\onlinecite{aulbur2000} & Ref.~\onlinecite{fleszar2001} \\
\hline
C   & 21.42 & 21.52  & 21.51 &  -- \\
Si  & 11.66 & 11.58  & 11.58 & 11.47 \\ 
Ge  & 12.44 & 12.48  & 12.46 & 12.43 \\
GaAs& 12.36 & 12.33  & 12.33 & 12.40 \\  
SiC & 15.25 & 15.23  & 15.23 & -- \\
GaN & 15.67 & 15.64  & 15.64 & -- \\ 
AlN & 14.83 & 14.86  & 14.85 & -- \\
\end{tabular}
\end{ruledtabular}}
\end{table}

\subsection{Potentials}
\label{sec_pot}

We now discuss the differences between the local KS potentials and densities obtained by
the LDA, EXX-OEP, and RPA-OEP methods.
For semiconductors such a comparison has been made before,\cite{godby1988,stadele1997,stadele1999,aulbur2000,betzinger2011}
and the main findings can be summarized as follows:
The EXX-OEP potential is more attractive than the LDA potential in the bonds but more repulsive in the interstitial regions.
The LDA potential is more spherically symmetric around the atoms, 
whereas the EXX-OEP potential shows more radial structure.
These differences in the potentials are reflected also in the density, for example, the EXX-OEP density
is larger in the bond regions; this can be attributed to the fact that EXX is one-electron self-interaction free.

\begin{figure}[h]
\centerline{\includegraphics[height=3.8cm]{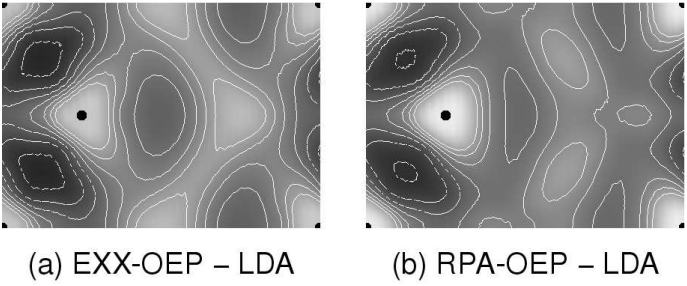}}
\caption{Difference of the EXX-OEP and LDA local potentials (left) and the RPA-OEP and LDA potentials (right)
of diamond in the (0$\bar1$1) plane.
Dark color represents negative values (EXX-OEP or RPA-OEP potential more attractive than LDA),
positive values are shown with light colors.
The contours are drawn for integer values of the difference (in eV). 
Black dots indicate the atomic positions. (Note that the potentials in OEP methods
are only determined up to a constant shift.)}
\label{fig_diam_2Dpot}
\end{figure}

\begin{figure}[h]
\centerline{\includegraphics[height=5.5cm]{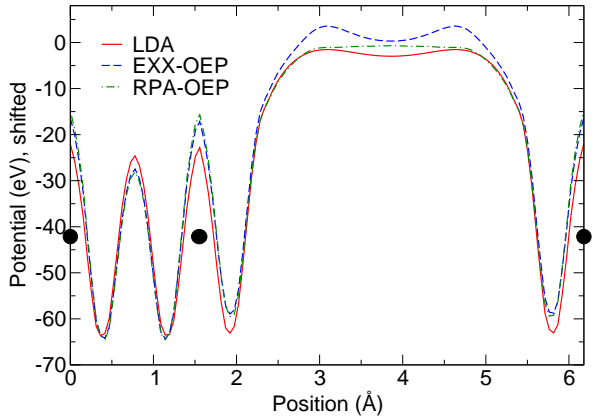}}
\caption{The local XC KS potential along the [111] direction in diamond 
as obtained with LDA, EXX-OEP, and RPA-OEP schemes. 
The potentials were shifted so that the values coincide at the intersection of the atomic PAW sphere
with the bond.}
\label{fig_diam_locpot}
\end{figure}

Our data share these main features discussed in previous publications.
For example, Fig.~\ref{fig_diam_2Dpot}(a) shows a slice of the difference between the EXX-OEP 
potential and LDA potential in the (0$\bar1$1) plane of diamond.
The darker regions indicate a deeper EXX-OEP potential, this can be seen for the bond region 
between two atoms (between the black dots).
The EXX-OEP potential is more repulsive in the interstitial regions, as shown by light gray
on the right hand side of Fig.~\ref{fig_diam_2Dpot}(a).
Figure~\ref{fig_diam_locpot} shows the potentials along the [111] direction (shifted to coincide at the intersection 
of the bond and the PAW atomic sphere).
One can see that LDA gives smaller difference than EXX-OEP in the potential between the bond and interstitial regions.
Moreover, the larger asymmetry of the EXX-OEP potential around the atoms compared to the LDA potential
can be clearly observed.

\begin{figure}[h]
\centerline{\includegraphics[height=5.5cm]{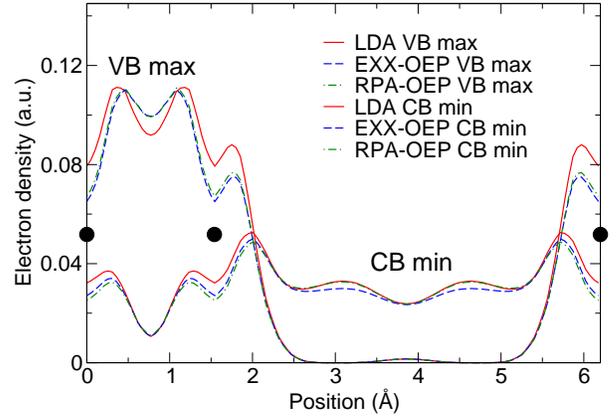}}
\caption{The electron charge density of the orbital at the valence band maximum (VB max, at $\Gamma$) and 
at the conduction band minimum (CB min, at X) of diamond along the [111] direction as obtained by LDA, EXX-OEP, and RPA-OEP.
}
\label{fig_diam_eldens}
\end{figure}

The RPA-OEP potentials and densities were compared to the LDA results by Godby~{\it et al}.\cite{godby1988}
Also there a deeper potential in the bonding region was observed and assumed to be the 
origin of the larger RPA-OEP gap compared to LDA.
The difference of the RPA-OEP and LDA potentials for diamond is shown in Fig.~\ref{fig_diam_2Dpot}(b)
and, on first sight, it is very similar to the difference between EXX-OEP and LDA.
However, there is a significant difference in the interstitial region 
where the RPA-OEP potential is more similar to the LDA potential.
In fact, as can be seen in Fig.~\ref{fig_diam_locpot}, the RPA-OEP potential varies very little
in the interstitial regions, especially compared to the EXX-OEP potential.
This can be attributed to the fact that the $GW$ self-energy includes also the response of the 
surrounding electron density (at the RPA level), not accounted for in the exchange-only 
case.\cite{eguiluz1992,niquet2003pra,niquet2003}
When the differences between the bonding and interstitial regions are compared, the EXX-OEP 
is clearly less attractive in the interstitial region.
As the interstitial regions are dominated by unoccupied orbitals, the higher EXX-OEP 
potential increases their energy relative to the occupied states, which dominate the bonding regions.
This effect increases the KS gap of the system with EXX-OEP compared to RPA-OEP.
The differences in the EXX-OEP and RPA-OEP potentials lead also to changes in the electronic states.
For example, as shown in Figure~\ref{fig_diam_eldens}, the more repulsive EXX-OEP potential
in the interstitial region (around 3~${\rm \AA}$ in Figure~\ref{fig_diam_eldens})
leads to a reduction of the charge density of the state at the conduction band minimum.
This increase of the EXX-OEP potential compared to RPA-OEP in the interstitial regions relative 
to the bonding regions is a general property and we observe it for other systems as well.

We note that the differences in the valence densities obtained from the LDA, EXX-OEP, and RPA-OEP 
potentials are in qualitative agreement with densities obtained by LDA, HF, Coulomb hole plus screened exchange (COHSEX), and scQP$GW$
in Ref.~\onlinecite{bruneval2006}.
Namely, the HF valence density is increased in the bonding region, as it is with EXX-OEP compared to LDA.
The RPA-OEP, COHSEX, and scQP$GW$ densities in that region are in between the other two approaches.
In the interstitial region, the HF density is reduced.
Also in our case, the EXX-OEP densities are very similar to the self-consistent HF densities, 
differing only around the cores because of the different one center terms.

\begin{figure}[h]
\centerline{\includegraphics[height=5.5cm]{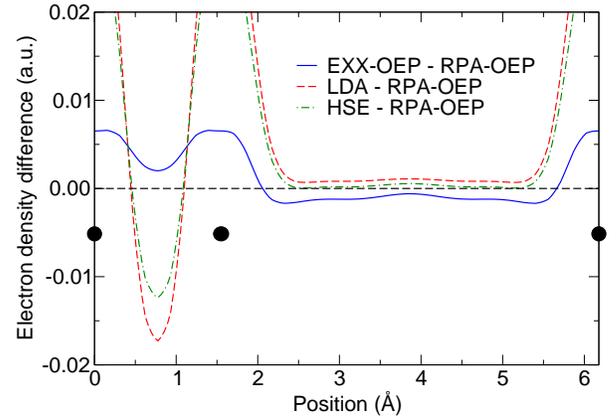}}
\caption{The difference between the EXX-OEP, LDA, and HSE total valence electron densities and the RPA-OEP valence electron density 
in diamond along the [111] direction.} 
\label{fig_c_densdiff}
\end{figure}

\begin{figure}[h]
\centerline{\includegraphics[height=5.5cm]{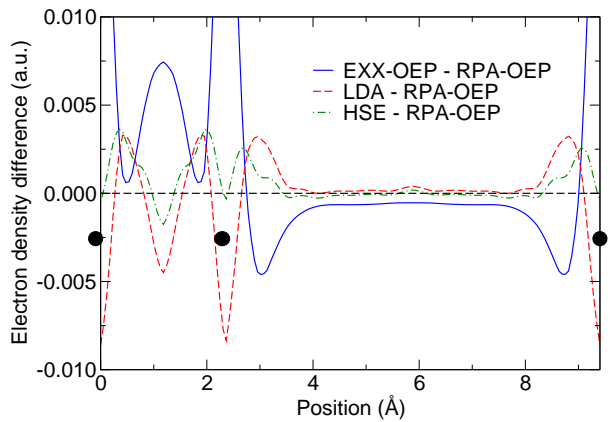}}
\caption{The difference between the EXX-OEP, LDA, and HSE total valence electron densities and the RPA-OEP valence electron density 
in bulk Si along the [111] direction.}
\label{fig_si_densdiff}
\end{figure}

It is interesting to ask how well do different functionals describe the electron density.
If we assume that the RPA-OEP density is close to the exact one, we can use it as a reference.
As stated before, the LDA predicts a density which is too low in the bonding region of, for example, diamond.
This is not unexpected since the strongly covalent character of bonding is far from the uniform electron gas limit
and self-interaction errors are present in the LDA.
The usual approach to alleviate some of the self-interaction error in the XC functional is to include a fraction 
of the non-local Hartree-Fock-like potential, as done, for example, in the Heyd, Scuseria, and Ernzerhof (HSE) functional.\cite{adamo1999,heyd2003hse,heyd2006hse}
However, the screening is quite weak in diamond and the RPA-OEP density is close to the EXX-OEP density in the
bonding region, as shown in Fig.~\ref{fig_c_densdiff}.
Including 25~\% of the exact-exchange, as done in the HSE functional, is not sufficient to bring the HSE density
much closer to the RPA-OEP density.
A different image appears for Si, where the screening is stronger, consequently the RPA-OEP and EXX-OEP
densities differ more in the bonding region, see Fig.~\ref{fig_si_densdiff}.
In this case the RPA-OEP density is closer to the LDA density than to the EXX-OEP one
and the recipe to use 25~\% of the exact-exchange is quite successful 
as the HSE density is close to the RPA-OEP one.
Interestingly, HSE agrees well with RPA-OEP for the description of the density in the interstitial regions of Si, C, and of other
systems, e.g., Ne (not shown).

\section{Discussion and conclusions}

We have presented results obtained within two OEP schemes, the EXX-OEP which
includes only the exchange diagram, and RPA-OEP (with the scQP$GW$ self-energy)
which also includes RPA screening.
Our main focus was to obtain the KS potential and ``band gaps",
both the single particle Kohn Sham gaps and the true gaps including the contribution of the derivative discontinuity.
Clearly, the contribution of the derivative discontinuity to the band gap is 
large, usually between 20--50~\% for the RPA functional,\cite{godby1988,gruning2006}
and therefore the KS single particle gap 
is too small compared to the QP gap observed experimentally.
As stressed before, the agreement of the EXX-OEP gaps
with experimental data is merely accidental and does not hold for systems with
an experimental gap above approximately 3.5~eV.\cite{magyar2004}
In fact, when the derivative discontinuity contribution is added to the EXX-OEP data,
the gap approaches the HF gap, known to be much too large.

\begin{figure}[h]
\centerline{\includegraphics[height=3.5cm]{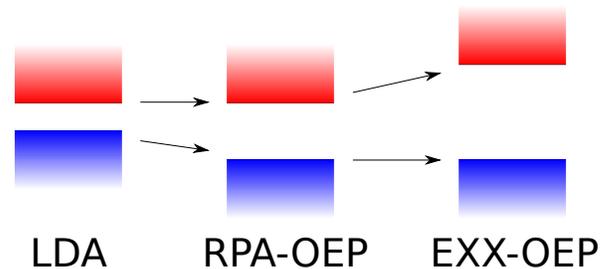}}
\caption{A schematic illustration of the increase of the KS gaps between the LDA, RPA-OEP, and EXX-OEP
approaches.
Blue and red denotes the valence and conduction band states, respectively.
}
\label{fig_gap_scheme}
\end{figure}

A clear trend in the KS single particle gaps is observed for all considered materials, 
the LDA gap is smaller than the RPA-OEP one, and the EXX-OEP gap is the largest.
This can be explained from the variations in the local potentials, and we
show this schematically in Figure~\ref{fig_gap_scheme}.
We note that the OEP potential can be obtained only up to a constant shift but 
the differences in the potentials clearly show more attractive and more repulsive regions. 
The LDA and RPA-OEP potentials are similar in regions of low density, the RPA-OEP potential, 
however, is more attractive in the bonding region where the density of the valence band states is large.
This lowers the position of the valence band states and increases the gap compared to LDA.
While the EXX-OEP and RPA-OEP potentials are similar in the bonding region, there are 
significant differences in the interstitial regions where RPA-OEP is more attractive.
As the interstitial regions are dominated by the conduction band states, 
the more repulsive EXX-OEP potential increases the energy of the conduction band states so that the EXX-OEP 
gap becomes even wider than the RPA-OEP gap.
The changes in the potentials are also reflected in the electron density, leading, e.g., to an
increase of the density of the diamond valence band maximum in the bonding region, shown in Fig.~\ref{fig_diam_eldens}.

The main subject of our work was the investigation of the true final quasi-particle gaps, which  can be compared to the
experimental fundamental gap, {\em i.e.} the difference between the experimental electron affinity and ionization potential.
To obtain the theoretical prediction of the gap, two methods can be used.
First, the standard Green's function approach, $G_0W_0$, where the self-energy correction to the gap is renormalized
by a scaling factor $Z$ that derives from the energy dependence of the self-energy. 
Second, one can calculate the contribution to the derivative discontinuity of the XC potential,
which in this case--- as derived by Niquet and Gonze ---differs from the $G_0W_0$ expression
 by the lack of this $Z$ renormalization factor.
We prefer the standard $G_0W_0$ expression where $Z$ is included and which thus takes into account 
the energy dependence of the self-energy.
Moreover, as discussed previously, the absence of the $Z$ factor is, strictly speaking, not justified,
as also demonstrated by the two plane wave model calculations of Lannoo~{\it et al.}\cite{lannoo1985} Furthermore,
the errors of the gaps are quite large when $Z$ is not included.

One can expect that the self-consistency introduced by the OEP procedure will lead to 
a more consistent starting point for the quasi-particle calculations, as compared to the 
standard GGA or LDA input.
This is indeed the case, and we obtain similar errors for the gaps for all the materials studied, including
problematic cases, such as ZnO and Ge.
The only major drawback is that 
the $G_0W_0$@RPA-OEP gaps are consistently larger than the experimental fundamental gaps.
At first sight, this is hardly satisfactory, since the computational cost of the method is 
about one order of magnitude larger than for standard $G_0W_0$@PBE calculations.
And $G_0W_0$@PBE calculations in many cases yield seemingly better agreement with experiment. 
When we compare to other computational approaches, we note that recent results for 
$GW^{\rm TC-TC}$@HSE calculations
(including a vertex in the screening) also yielded
consistently too large band gaps that are, in fact, within about 0.1~eV of the present values.\cite{grueneis2014}
Compared to the  $GW^{\rm TC-TC}$@HSE  the present approach is fairly cheap:
evaluation of the electron-hole interaction in the screening $W^{\rm TC-TC}$ requires solving the Bethe-Salpeter
equation, an order $N^5$ step. 
This is prohibitively expensive for large systems. 
Viewed from that perspective, 
the RPA-OEP method  seems to present a very convenient shortcut and certainly provides
an excellent and concise starting point for subsequent $GW$ or RPA total energy calculations, without the need for the 
expensive $N^5$ Bethe-Salpeter step.

Since two quite different theoretical approaches yield consistently too large gaps,
one might ask whether the present values are in fact not accurate.
While there will be some contribution from electron correlation diagrams not accounted for
here or in Ref.~\onlinecite{grueneis2014}, e.g. electron-electron or hole-hole ladders, the residual error might be 
related to the neglect of the electron-phonon part of the electron self-energy, which  was not considered in this work
or Ref.~\onlinecite{grueneis2014}. 
In fact, it has been shown that the electron-phonon interactions can affect the value of the gap significantly; for example, a reduction
of $\approx$0.4~eV was obtained recently for the direct band gap of diamond.\cite{giustino2010,cannuccia2011,ponce2014}
Since our direct gap in diamond is also $\approx$0.4~eV larger than the experimental value, accounting for electron-phonon
interactions would lead to a better agreement with experiment.
Interestingly, also the gaps of ionic materials, in particular LiF, are 
overestimated in our calculations, which might also be related to 
 the electron-phonon contributions  (see also Ref.~\onlinecite{botti2013}).
Thus to make a final judgment of the quality of the band gaps of the $G_0W_0$@RPA-OEP scheme,
electron-phonon interactions should be accounted for.
This is, however, beyond the scope of the present work.

\begin{acknowledgments}

This work was supported by the Austrian Science Fund (FWF)  within the SFB ViCoM (Grant F 41).

\end{acknowledgments}


\begin{thebibliography}{100}
\bibitem{footGKS}%
  \BibitemOpen
  \bibinfo {note} {Actually, in the case of hybrid functionals a non-local
  potential is used corresponding to the generalized Kohn-Sham scheme.}%
  \bibAnnoteFile{Stop}{footGKS}%
\bibitem{note_gaps_increase}%
  \BibitemOpen
  \bibinfo {note} {Although we observed an increase of the transition energies
  for most of the systems considered in this study, it is not a general rule.
  For example, in Ge when the cut-off energy is increased the
  $\Gamma$$\rightarrow$$\Gamma$ and, to a smaller extent,
  $\Gamma$$\rightarrow$L transition energies decrease, the
  $\Gamma$$\rightarrow$X energy difference increases.}%
  \bibAnnoteFile{Stop}{note_gaps_increase}%
\end{thebibliography}

%
\end{document}